# Single-photon-emitting diodes: a review

**A. J. Bennett**[*,1], **P. Atkinson**[2], **P. See**[1], **M. B. Ward**[1], **R. M. Stevenson**[1], **Z. L. Yuan**[1], **D. C. Unitt**[1,2], **D. J. P. Ellis**[1,2], **K. Cooper**[2], **D. A. Ritchie**[2] and **A. J. Shields**[1]

[1] Toshiba Research Europe Limited, Cambridge Research Laboratory, 260 Science Park, Milton Road, Cambridge, CB4 0WE, United Kingdom

[2] Cavendish Laboratory, Cambridge University, J. J. Thomson Avenue, Cambridge, CB3 0HE, United Kingdom

## Abstract

Compact and reliable sources of non-classical light could find many applications in emerging technologies such as quantum cryptography, quantum imaging and also in fundamental tests of quantum physics. Single self-assembled quantum dots have been widely studied for this reason, but the vast majority of reported work has been limited to optically excited sources. Here we discuss the progress made so far, and prospects for, electrically driven single-photon-emitting diodes (SPEDs).

[*] Corresponding author: e-mail: anthony.bennett@crl.toshiba.co.uk, Phone: +44 (0) 1223 436941, Fax: +44 (0) 1223 436909

# Introduction

Self-assembled InGaAs/GaAs quantum dots (QDs) have been widely studied since the original suggestion that they may offer some advantages over quantum-wells was first made in 1981 [1]. Arguably, the most promising application for quantum dots now is in single-photon generation. Many fundamental tests of quantum mechanics can be made with single photons but recently interest in these sources has blossomed due to the discovery of protocols for handling quantum information with light, either within a photonic quantum computer [2] or with quantum cryptography [3]. In the near to medium term, a single photon source is likely to find its first practical application in a quantum cryptography system in which it can improve security and efficiency.

The majority of the reported work on single photon generation from single quantum dots involves optical excitation. For practical reasons, if QD-based single-photon sources are ever to achieve wider application outside the laboratory it would be beneficial if they were electrically-driven. Ideally, such devices would be small, robust, efficient and have a low probability of emitting more than one single photon. A series of single-photon-emitting diodes (SPEDs) have now been reported [4, 5, 6]. These structures are particularly promising as all the tricks of the conventional optoelectronics industry, such as band-gap engineering, photonic confinement and on-chip integration, to name a few, can be brought to bear on their development. Already significant progress has been made towards these aims and it is the object of this paper to summarise these results and outline the way forward.

# A simple SPED

Much of the pioneering work on optical spectroscopy of single quantum dots was performed by Gerhard Abstreiter's group during the early 1990's [7]. In 2000, a fundamental experiment was reported showing that single photons could be generated from single self-assembled QDs [8]. This Hanbury-Brown and Twiss experiment [9] is conceptually very simple, consisting of only a beam-splitter, two photodiodes and some electronics (Fig. 2 (b)). Over time, a histogram is built up of number of the events where both detectors count a photon as a function of the time between the two detection events. Put simply, a single photon cannot be detected at the same time at both detectors and therefore an absence of cross-counts in the histogram is seen near time zero. This histogram is a direct measurement of the auto-correlation function, $g^{(2)}(\tau)$, of the light field.

Similarly, electrically injected devices based upon single QDs were reported as early as 2000, [10, 11] and it did not take long for a Hanbury-Brown and Twiss auto-correlation measurement on a single-photon-emitting diode to be reported [4]. This simple SPED consisted of a vertical *p-i-n* junction with a layer of low density quantum dots within the intrinsic region. An opaque metallic film patterned with micron-sized apertures on the surface of the device allowed emission to be collected from only one dot (Fig. 1 (a)). It was shown that by increasing the voltage across the contacts to above ~1.5V carriers could be injected into the device and some would result in emission from the dot directly beneath the aperture. Auto-correlations reported on this device under DC electrical injection displayed a reduction in the number of coincidences observed on the two detectors at time zero. In addition, a milestone result was reported showing the strong suppression of multi-photon emission using pulsed electrical injection. It was found that the use of pulses nominally 400ps in duration was sufficient to observe $g^{(2)}(0) \sim 0.11$ [4].

In general, several factors determine the $g^{(2)}(0)$ that is observed:

[i] There is always background emission picked up by the detectors, which results in a photon count rate $R_{BK}$. This can be due to multi-photon emission from other states in the source, uncorrelated emission from other states in the same or other QDs, or merely stray light.

[ii] There are also dark counts associated with each detector, resulting in a count rate $R_D$, which occurs when the detector fires in the absence of a photon.

If the signal from the perfect single photon source results in a count rate, $R_S$, and both detectors detect contributions equally then it can be shown that:

$$g^{(2)}(0) = \frac{2(R_D + R_{BK})R_S + (R_D + R_{BK})^2}{(R_D + R_{BK} + R_S)^2}$$

(1)

[iii] Finally, the pulse length, $\Delta t$, must be small relative to the radiative lifetime, $\tau_{rad}$, of the state being studied so refilling cannot occur. Even for pulses nominally 400 ps long this refilling was not an issue for the original device, where all multi-photon counts could be attributed to detector dark counts and background light.

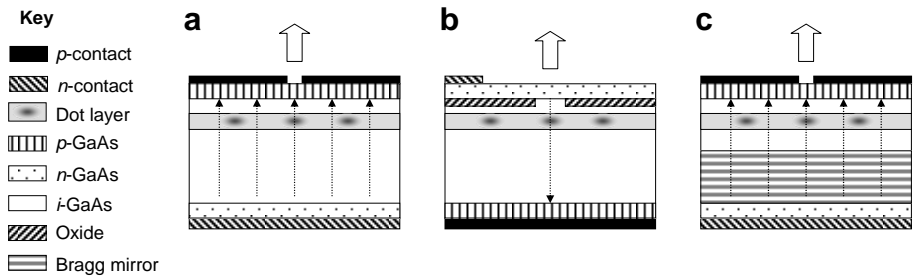

**Fig. 1** Three types of single photon emitting diode (SPED). (a) The first SPED that was reported [4], (b) the "oxide-aperture SPED" [5] and (c) the "planar-microcavity SPED" [6].

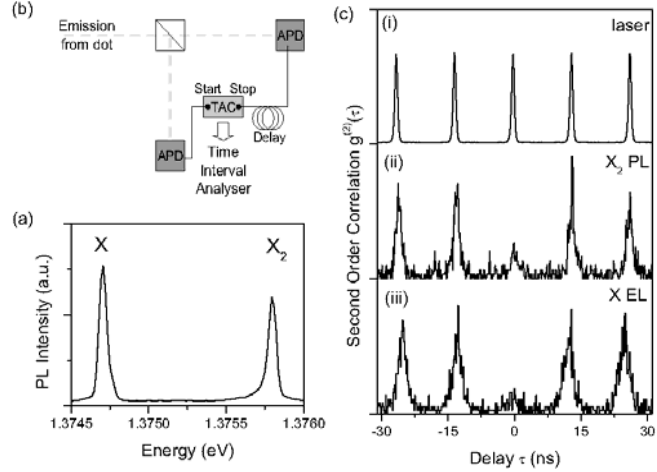

**Fig. 2** (a) Emission spectra from the device reported by Yuan *et al* [4]. (b) Schematic diagram of the Hanbury-Brown and Twiss interferometer. (c) (i) Auto-correlation data from an 80 MHz pulsed laser, (ii) the device under pulsed electrical emission when centered on the (i) wetting layer, (ii) $X_2$ state and (iii) X state.

## An oxide-aperture SPED

In this initial single photon source currents of up to 6 μA were required to saturate the exciton ($X$) emission data under DC voltage [4]. Assuming an exciton state lifetime of ~ 1ns this corresponds to > $10^4$ electrons per photon emission event. This is not surprising as the device measured several tens of microns in each direction and the capture cross-section of the QD was significantly smaller. Whilst this was not a problem in terms of performing auto-correlation measurements we can envisage several interesting experiments where only a single dot is excited. A few publications have now appeared towards this goal [5, 12, 13].

The approach we have taken builds on a device design that has been used to demonstrate ultra-low threshold VCSELs [14]. In this second structure (Fig. 1 (b)) an aluminium arsenide layer, which was oxidised in a humid atmosphere, forms an insulating aluminium-oxide annulus within the device with a small unoxidised aperture in the center, where a QD was located. Voltage applied to the device was only dropped across this central region and current only injected into the micron-sized aperture. We were then able to saturate the emission from the $X$ state with currents of ~ 10 nA, corresponding to tens of carriers being injected into the device for every single photon emitted.

The turn-on voltage ($V_T$) for these devices was ~ 2.0 - 2.2 V due to the resistance of the ohmic contacts used. The dimensions of the aperture could be estimated by driving the device at a large bias above threshold. Bright electroluminescence from the GaAs band-edge states could be seen to only occur from a ~ 2 x 2 μm region within the center of the 30 x 30 μm mesa (Fig. 3). In fact, spreading of the current as it passes though the area was significant under these conditions. We have found that at lower voltages, when emission from the QD was dominant, the area injected with carriers was actually less than 1 x 1 μm [5].

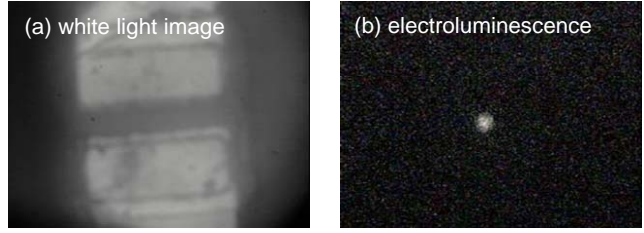

**Fig. 3** (a) White light image of the 30 x 30 μm mesa. (b) Image of the electroluminescence of the device at high bias (light emission is occurring from the GaAs band-edge in the structure).

Electroluminescence spectra recorded from this single QD at 2.3 V are shown in Fig. 4 (a). Through power dependent data (Fig. 4 (b)) and study of the QD emission lines as a function of voltage we have been able to assign the two most prominent emission lines in Fig. 4 (a) to $X^+$ and $X_2^+$ states of the QD [5]. The fact that voltages > 2V were required to inject current into the device means that a finite electric field across the intrinsic region favours an imbalance of charge on the dot.

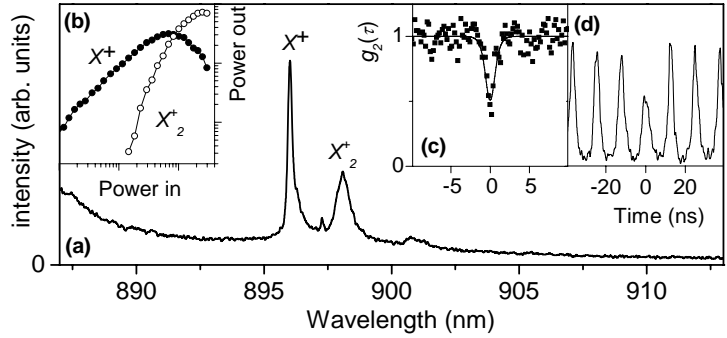

**Fig. 4** (a) Electroluminescence spectrum recorded from the oxide-aperture SPED at 2.30V. (b) Power dependent electroluminescence from the $X^+$ and $X_2^+$ states. (c) and (d) show auto-correlation histograms from the $X^+$ state under DC and pulsed electrical injection, respectively.

Fig. 4 (c) shows an auto-correlation histogram obtained under DC electrical injection for the $X^+$ state. We observed a clear dip at zero time delay, indicating anti-bunching of the emitted photons. Modelling this data suggested the $X^+$ state had a radiative lifetime of 450 ps.

We repeated this experiment using electrical pulses of nominal length 300 ps and at a repetition rate of 80 MHz, to drive the device with a time-averaged current of 10 nA (Fig. 4 (d)). The reduction in area of the zero-delay peak clearly indicates non-classical emission. Spectral measurements suggest the area of central peak should be 24% if the signal was due entirely to a perfect single-photon-emitting emission line upon a spectrally flat background emission with classical photon statistics. As discussed earlier, the remaining area of the peak at time zero can be attributed to the refilling of the dot after the emission of a first photon.

## The photon collection efficiency from a planar microcavity

In the previous devices (Fig. 1 (a) and (b)) the high refractive index of GaAs ($n_{GaAs}$ ~ 3.5 at 900 nm [15]) leads to a low proportion of light escaping from the top surface of the crystal. An analytic expression for the collection efficiency can be derived using the fact that the excitonic ground state of these quantum dots consists of two orthogonal dipoles orientated in the plane of the crystal. In addition, reflection from the air/GaAs interface further reduces the collected fraction. A lens of numerical aperture (NA) 0.5 can only collect light hitting the air-GaAs interface at angles less than $\sin^{-1}(NA/n_{GaAs})$ which corresponds to a collection efficiency, $\eta$ [16]:

$$\eta = \left(1 - \left(\frac{n_{GaAs}-1}{n_{GaAs}+1}\right)^2\right) \cdot \left(\frac{1}{2} - \frac{3}{8} \cdot \cos(\sin^{-1}(\frac{NA}{n_{GaAs}})) - \frac{1}{8} \cdot \cos^3(\sin^{-1}(\frac{NA}{n_{GaAs}}))\right) = 0.5\% \quad (2)$$

It has long been known that modifying the structure around a light emitting active region can increase this collection efficiency. Interface surface roughness, solid immersion lenses and microcavities have all been considered [17, 18, 19, 20]. With broadband light sources, such as quantum wells, optimisation of these structures is non-trivial due to the wavelength and position dependence of the cavity. Consequently any modelling of the structure must consider all wavelengths of emission at all positions in the cavity and average these results. In contrast, single quantum dots have narrow emission lines and are located at fixed positions, so electromagnetic modelling is simpler.

Planar microcavities can offer significant enhancements in collection efficiency. They are also particularly easy to implement. Therefore, we have undertaken a theoretical study of the photon collection efficiency for various designs of planar microcavities using the CAvity Modelling FRamework Package (CAMFR: from http://camfr.sourceforge.net/) to optimise our cavity design. The structures we have considered had a single dipole orientated parallel to the cavity emitting at a wavelength coincident with the cavity design wavelength. The Poynting vector (a measure of power flow through an area) was then calculated over the surface of a sphere of radius 10 µm, centered on the dipole. This sphere radius is many times the wavelength of the emitted radiation and therefore is a fair approximation to the far field emission pattern of the dipole-cavity system. The results of this calculation were then used to determine the total power flow at an angle $\theta$ to the growth direction. The collection efficiency is then the sum of the power into a certain numerical aperture ($\theta = 0$ to $30°$ for NA = 0.5) divided by the sum of the power emitted at all angles ($\theta = 0$ to $180°$).

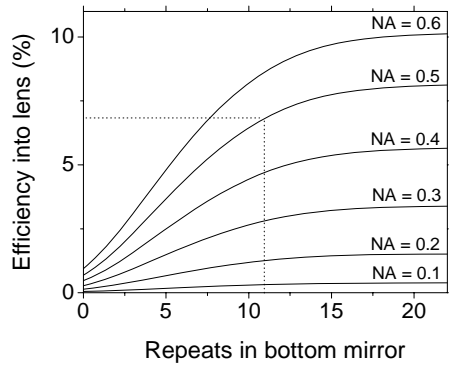

**Fig. 5** Modelled collection efficiencies for light emitted from a dipole 2 wavelengths below the GaAs/air interface and 1 wavelength above a GaAs/AlAs Bragg mirror as the number of repeats in this mirror is increased.

Calculations were performed of the collection efficiencies for light emitted from a dipole two wavelengths below the GaAs/air interface and one wavelength above a GaAs/AlAs Bragg mirror as the number of repeats in this mirror is increased. This was carried out for a series of different numerical aperture lenses (Fig. 5). The results show that a maximum efficiency would be observed for the largest number of periods on the bottom mirror. This efficiency tends asymptotically to approximately 8 % with the number of repeats for a NA = 0.5 lens.

We can now look more closely at the emission pattern from the structure with 12 mirror repeats (Fig. 6 (b)) and the emission from a comparable structure without the Bragg mirror (Fig. 6 (a)). When there is no cavity interference occurs between the downward travelling light and light that is emitted upward and then reflected from the air/GaAs interface, resulting in a periodic variation in intensity with angle between 90 and 180°. Calculating the proportion of light that is emitted into an NA of 0.5 we obtain an efficiency of 0.5%, as expected from equation (2). In contrast, in the microcavity sample there is a strong suppression in the emission directly downwards due to the high reflectivity of the Bragg mirror at normal incidence. A significant increase in the power flow at 90° is also observed because this structure is an effective waveguide. The model predicts that it should be possible to collect 7.0% of the radiation emitted by this structure.

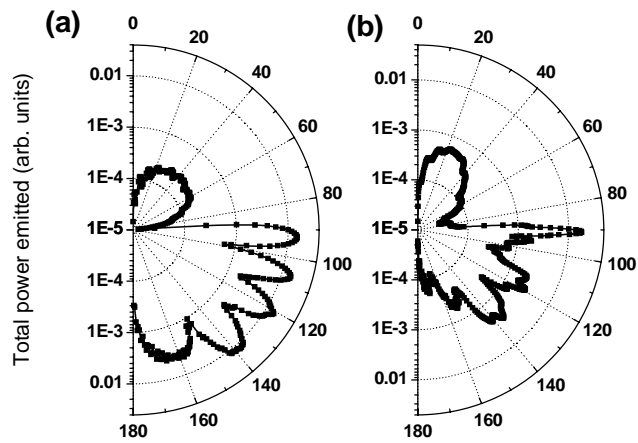

**Fig. 6** Modelled emission patterns of single photon sources with (a) dots two wavelengths below the surface with no cavity and (b) with a 3-wavelength cavity upon a 12 period GaAs/AlAs mirror, again with the dot two wavelength below the upper surface. In both cases the total power emitted at a given angle is plotted. Therefore the collection efficiency into an NA = 0.5 lens (corresponding to an angle of 30°) is given by the integral of the power emitted into 0 to 30° and divided by the total emitted power from 0 to 180°. For case (a) this is 0.5%, for case (b) 7.0%.

We have also considered if any further improvement in collection efficiency can be gained from including an upper GaAs/AlGaAs Bragg mirror in the planar cavity. There are two factors at play in this series of results: firstly, as the number of periods in the top mirror is increased the emission is better collimated in the upward direction and the collection efficiency increases. However, beyond a certain point the collection efficiency then decreases as the upper mirror directs more of the radiation downwards. We have found that for a one-wavelength cavity with a 12-period bottom mirror the optimum number of top mirror periods is 4, increasing the collection efficiency to 11.8%. This represents a ~ 24-fold improvement in collection efficiency relative to a sample without a microcavity.

### A microcavity SPED

We now discuss a SPED that incorporates this cavity. The device is shown schematically in Fig. 1 (c). The QD layer is located one wavelength above a *n*-doped 12 period GaAs/Al$_{0.98}$Ga$_{0.02}$As Bragg mirror and two wavelengths below the air/GaAs interface.

In the best device found, study of the saturated *X* and *X$_2$* intensities allowed us to experimentally measure the photon collection efficiency into our NA = 0.5 lens to be 4.7 ± 0.5 % [6]. This is a ten-fold improvement of the maximum possible collection efficiencies that could be observed with the previous device designs. However, it falls short of the 7.0 % efficiency predicted in Fig. 5 which may be due to a non-optimal alignment of the QD beneath the aperture in the metallic surface layer. Nevertheless, due to the improved photon collection efficiency in this sample it was possible to more closely investigate the photon emission statistics of the states by performing correlation measurements in tens of minutes whilst obtaining good signal-to-noise ratios.

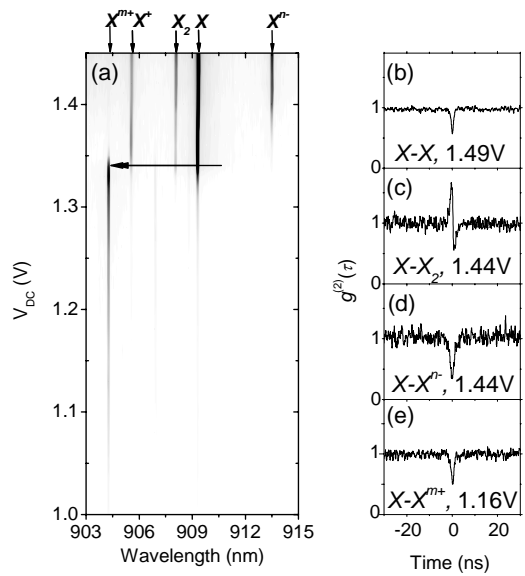

**Fig. 7** (a) Photo-luminescence spectra as a function of voltage for the sample under weak optical excitation. (b) Auto-correlation measurement on the *X* state under DC electrical injection at 1.49V. Cross-correlation measurements are then shown for (c) the *X* and *X$_2$* states at 1.44V, (d) the *X* and *X$^{n-}$* states at 1.44V and (e) the *X* and *X$^{m+}$* states at 1.16V .

Measurement of the emission from this source as a function of voltage revealed a number of emission lines (Fig. 7 (a)). In order to confirm that these were all due to emission from the same QD, cross-correlation measurements were recorded. In this experiment, rather than the two detectors in the Hanbury-Brown and Twiss experiment detecting photons from the same light field, one detector collects photons emitted from one state and the other from a different state. The results of these cross-correlations are shown in Fig. 7 (b)-(e). Those recorded at voltages greater than 1.45V employed a DC voltage to inject carriers whilst those recorded with voltages less than 1.45V used optical excitation with continuous-wave laser. In Fig. 7 (c), a cross-correlation measurement on the two emission lines at 909.3 and 908.3 nm confirms these lines originate from a photon cascade, supporting our assignment of these states as exciton (*X*) and biexciton (*X$_2$*). Cross-correlations between the *X* state and the lines at 913.5 and 904.1 nm (Fig. 7 (d) and (e), respectively) both show anti-correlation suggesting they originate from mutually exclusive complexes within the same QD. This supports our conclusion [21] that these states are due to multiply positively and negatively charged exciton states, *X$^{n-}$* and *X$^{m+}$*.

Auto-correlation data was also recorded with the device driven in a pulsed electrical mode. A fixed bias just below the turn-on voltage was applied to the device and nominally 300 ps long voltage pulses were used to inject carriers into the optically active region at a repetition rate of 80 MHz. Once again, clear suppression of the central peak area for both the $X$ and $X_2$ states indicated single photon emission was occurring from these states [6].

## Controlling the photon emission time and statistics

In optical excited auto-correlation experiments the repetition rate has been limited by the 80MHz operating frequency of the commonly used titanium-sapphire laser. With electrically driven sources we do not have this limit to the maximum operating frequency of the experiment. If the same electrical excitation scheme as has been previously discussed is used, biasing the device just below threshold ($V_{DC}$ < 1.45 V) and using low amplitude voltage pulses (height $V_{pulse}$) to inject carriers, then the radiative lifetime of the emitting state limits the repetition rate. We have found that frequencies of a few hundred MHz are achievable, at which point the visibility of the peaks in the auto-correlation data is low [22].

Shown in Fig. 8 are two time-resolved traces recorded from one of the microcavity SPEDs operating in pulsed mode at 80 MHz. The data in the upper panel was recorded in the "conventional" mode described previously. After the electrical pulse injects carriers into the dot the photon generation probability fell exponentially with a lifetime of 2.1 ns, which is the radiative lifetime of the $X$ state.

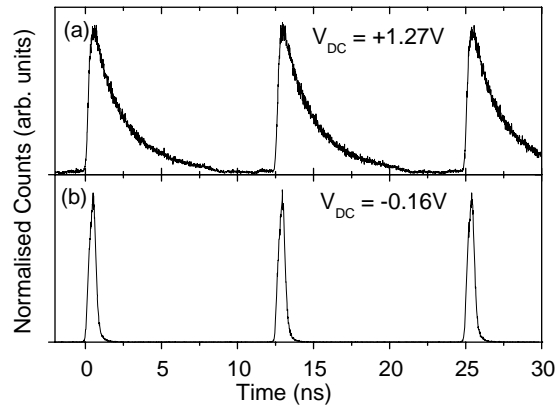

**Fig. 8** Time resolved data from the microcavity SPED using pulsed electrical injection with 400ps pulses at 80 MHz for two different $V_{DC}$ levels and the same current.

However, when the DC bias, $V_{DC}$, was reduced, and the pulse amplitude, $V_{pulse}$, increased to maintain the same time-averaged current flow, the jitter on the photon emission time fell drastically. When $V_{DC}$ = -0.16 V (Fig. 8 (b)) the decay time is limited by the timing resolution of our detection equipment. Similar reductions in the lifetime are observed for the $X_2$ state.

This drastic reduction in the apparent decay time occurs because when the pulse ends a large electric field across the device removes carriers from the QD, preventing further emission. We have found that the electric field at which this occurs for the particular QDs studied here (emitting in the range 900 - 920 nm) is ~1500 Vcm$^{-1}$. From spectroscopic studies of these QDs we find that it is only the electrons that are removed from the dot at electric fields larger than 1500 Vcm$^{-1}$ and that holes can remain on the dot until higher voltages are reached (this will be discussed later in this section) [21].

Operation of the device in the low timing-jitter mode obviously reduces the internal quantum efficiency as not all excitons are given time to radiatively decay. For the $X$ state studied here we have measured a reduction in the quantum efficiency by a factor of 5. The reduction is less for states with shorter radiative lifetimes, such as the $X_2$. Consider a comparison with the "simple" SPED described earlier [4]: the microcavity SPED has a photon collection efficiency 10 times larger, in the case of the $X_2$ the reduced jitter scheme lowers the internal quantum efficiency by a factor of two. However, as we shall now show, we can operate this source at a repletion rate of 1.07 GHz, 13.4 times faster than the 80 MHz discussed by Yuan *et al* [4]. The net result is that this source is capable of generating 65 times more single photons per second than the first SPED that was reported.

Fig. 9 (a) shows the auto-correlation trace we would expect if we had a classical jitter-free photon source operating at 1.07 GHz: here the width of the peaks are limited by the resolution of our detection system. However, for a source emitting single photons both detectors cannot measure the same photon at the same time, and hence the histogram will have a peak missing at time-zero (Fig. 9 (b)). Experimental data recorded from the $X$ and $X_2$ states at 1.07GHz is shown in Fig. 9 (c) and (d), respectively. A clear suppression of the area of the central peak in both cases indicates that single photon emission is occurring. It is noticeable that at low currents the finite delay peaks in the $X$ correlations display "anti-bunching". In other words, there is a reduced probability of the source emitting two single photons within a few nanoseconds. This is not observed for the $X_2$ state.

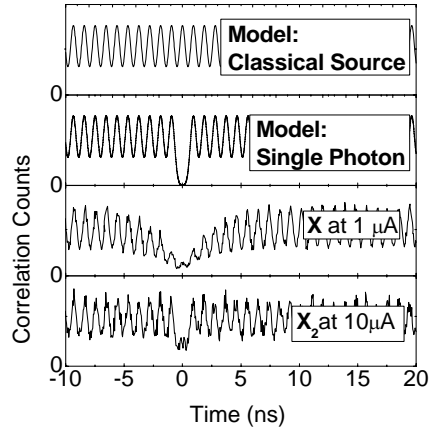

**Fig. 9** Auto-correlation data recorded at 1.07 GHz

Anti-bunching occurs because for non-resonant excitation (all electrical injection described here is non-resonant) the QD has a finite probability of being in a charged or dark state between consecutive pulses [23]. We have investigated the effect of different heights and shapes of voltage pulse in controlling the timescale of the anti-bunching. Our conclusion is that the timescale of the anti-bunching can be reduced by lowering the voltage for some or all of the time between pulses. Shown in Fig. 10 are data from auto-correlation traces recorded at 500 MHz as the $V_{DC}$ level between pulses was varied for the same current. The plots show the areas of each peak in the auto-correlation trace as a function of the time that peak is centered on. In Fig. 10 (b), where $V_{DC} = + 0.87$ V the anti-bunching time scale is several nanoseconds. However, when $V_{DC} = - 0.10$ V there is a clear reduction in the timescale of the anti-bunching effect. This shows that if large enough electric fields can be applied across the dot between injection pulses both electrons and holes will be removed, leaving the dot in the same state (empty) regardless of whether a photon was just emitted.

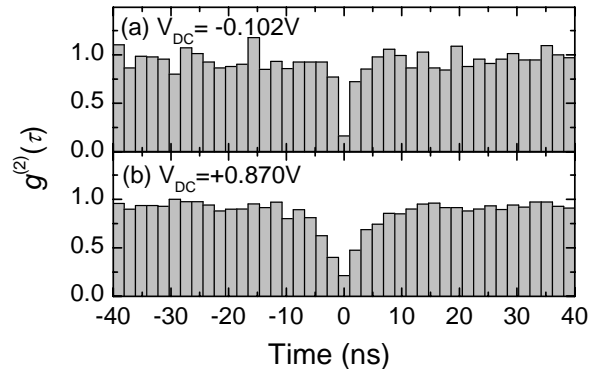

**Fig. 10** Plots of the peak areas in auto-correlation histograms recorded from the *X* state at 500 MHz and current 0.5 μA. At a $V_{DC}$ level of 0.870V anti-bunching of the photons ensures several peaks around the central peak at time zero are suppressed. As the $V_{DC}$ level is reduced this effect is suppressed.

## Future directions

To date, only a small number of SPEDs have been realised. At this early stage, we think it would be useful to specify a `wish list` of significant results that we believe could be reported over the next few years, some of which will be essential for the eventual commercial exploitation of these devices. We note that many of these advances would constitute significant advances whether seen with optical or electrically excited sources.

[i] The efficiency with which single photons are collected must be increased. The obvious solution here is to employ high quality-factor, low volume 3D microcavities which lead to a Purcell Effect [24, 25]. In this manner efficiencies of > 10% should be achievable within the near term. However, the problem of providing robust and reliable electrical injection to such cavities remains a challenging issue. In addition, it is clear that any reduction in radiative lifetime will increase the probability of a QD refilling within a given pulse and result in multiphoton emission.

[ii] The yield of useful devices must be increased. Growth of site-controlled QDs with a well defined emission wavelengths is most promising in this respect [26].

[iii] Controllable wavelength of emission. This would allow the wavelength of emission to be tuned to that of a cavity, to that of a second QD or another photon source.

[iv] For quantum cryptography telecoms wavelength single photon emission must be achieved. Rapid progress is being made in this area [13, 27, 28].

[v] Resonant tunnelling of carriers into a single QD, although not essential would potentially allow for a series of new and exciting experiments to be performed [29].

[vi] Higher temperature operation must be reported. At present all devices described here operate at temperature of 4-20K. In order for practical applications to be realised operation at 77K would be prefered.

[vii] For linear optics quantum computing single-photon or sub-poissonian emission is not sufficient. The photons that are emitted must also be time-bandwidth limited (i.e. indistinguishable in their time of emission and energy) [30].

[viii] The recent observation of optically-excited generation of entangled photon pairs from the excition cascade in a QD raises the possibility of exploiting this development in an electrically driven source [31].

## Conclusion

In conclusion, we have discussed the progress thus far, and future prospects for, single photon emitting diodes (SPEDs) based on single self-assembled InGaAs/GaAs QDs. It is clear that at present few technical barriers stand in the way of further development of these promising devices.

## Acknowledgements

DCU and DJPE would like to thank EPSRC and Toshiba for funding. This work was partially supported by the European Commission under Framework Package 6 Network of Excellence SANDiE and Integrated Projects through Qubit Applications (QAP, contract number 01584) and RAMBOQ (IST-2001-38864).